\documentclass[aps,prd,superscriptaddress,twocolumn]{revtex4}
\usepackage{amsmath,epsfig,mathrsfs,graphicx,amssymb,color}
\usepackage[T1]{fontenc}
\usepackage{t1enc}
\usepackage{hyperref}
\def\be#1\ee{\begin{equation}#1\end{equation}}
\newcommand{\ba}{\begin{eqnarray} }
\newcommand{\ea}{\end{eqnarray} }

\begin{document}

\title{General quantum measurements in relativistic quantum field theory}
\author{Adam Bednorz}
\affiliation{Faculty of Physics, University of Warsaw, ul. Pasteura 5, PL02-093 Warsaw, Poland}
\email{Adam.Bednorz@fuw.edu.pl}

\date{\today}

\begin{abstract}
Single particle detection is described in a limited way by simple models of measurements in quantum field theory. 
We show that a general approach, using Kraus operators in spacetime constructed from natural combinations of fields,
leads to an efficient model of a single particle detector. The model is free from any auxiliary objects as it is defined solely within the existing
quantum field framework. It can be applied to a large family of setups where the time resolution of the measurement is relevant,
such as Bell correlations or sequential measurement. We also discuss the limitations and working regimes of the model.
\end{abstract}
\maketitle

\section{Introduction}

Quantum field theory makes predictions about scattering and decays of particles that can be measured.
In contrast to quantum optics and condensed matter, where low energies allow very efficient detectors, high energy experiments
especially in the ultrarelativistic limit, cope with practical limitations, e.g. not all particles are detected
\cite{pdg}. 

The textbook approach is based on the relation between the transition probability between incoming and outgoing particles
and scattering matrix \cite{peskin}. This is completely nonlocal as the states are considered in momentum representation.
Exact modeling of real detectors is impractical because of enormous
technical complexity. Instead, simplified models such as the Unruh-DeWitt model have been proposed, involving an auxiliary particle traveling on its worldline, imitating a detector
\cite{unruh,witt,rsg,rove,lin,costa,martinez,anas,brody,gale}, reducing the notion of the particle to what a particle detector detects \cite{scul}.
Such models were fine in the early days of high energy physics when the actual low efficiency was not a particular problem. It was sufficient
 to map the collection statistics to the theoretical predictions about scattering and decays. 
 However, the Unruh-DeWitt and all existing constructions are unable to model a $\sim 100\%$ efficient detection of a single particle
 as a click \cite{click}. The real measurements in modern experiments have also to be localized in time and space. 
First, Bell-type tests of nonlocality \cite{epr,bell,chsh,chi,eber} require time resolution and high efficiency, achieved in low energy optics
\cite{hensen,nist,vien,munch}.
Higher energy attempts to make similar tests have failed so far \cite{belh}.
 Second, sequential measurements cannot destroy the particle after detection, allowing one to measure it once again. 
This kind of measurement is useful to reveal incompatibility between the measured quantities 
\cite{incom}.  It is possible for
immobile solid state objects \cite{solid} and only recently for photons
\cite{phot}. The high energy analogs are still awaited. 
In some cases, one can use the spacelike momentum to suppress the vacuum noise \cite{bb22} but it does not help to increase the efficiency.

In this work, we propose a different approach to measurement models in quantum field theory. Instead of auxiliary particles,
we shall directly define measurement Kraus operators \cite{kraus}, an element of positive operator-valued measure (POVM) \cite{peres,breuer},
within the existing  quantum field Hilbert space,
replacing the old fashioned projection, either in a direct manner or in the auxiliary detector's space. 
The Kraus operators are functionals of already existing fields. We show that a properly defined functional
can correctly represent the measurement. However, not all classes of such operators can serve as single particle detection models. 
To detect single particles (in a clickwise fashion), 
a nonlinear functional is necessary and it has a limited efficiency outside the safe energy regime. The best option is a universal measurement
of energy-momentum density, which models an almost perfect measurement in a wide range of parameters. 
The model is able to map the outcomes onto almost dichotomic events, with well separated absence and presence of the particle, when
a continuous flux of incoming particles arrives at the detector. Although the model is in principle perturbative in the detection strength, 
we are able to control potential higher order deviation and identify the working regime.

The paper is organized as follows. We start from the standard definition of generalized measurement operators, mapping them onto the quantum field theory framework.
Next, we present several natural classes of such a measurement, pointing out their weaknesses and advantages. Finally,
we discuss the applicability of the models to the Bell test and sequential measurements and confront them with other options like Unruh-DeWitt model. Lengthy calculations are left in the appendices.

\section{General quantum measurements}

From the obsolete projections, through auxiliary detectors, modern quantum measurements evolved to a description that requires no
extra objects. They are defined within the system's space by a set of Kraus operators $\hat{K}$ such that $\sum \hat{K}^\dag\hat{K}=\hat{1}$
and the state $\hat{\rho}$ after the measurement is transformed into $\hat{\rho}'=\hat{K}\hat{\rho}\hat{K}^\dag$ (no longer normalized)
with the probability $\mathrm{Tr}\,\hat{\rho}'$  \cite{kraus,peres,breuer}. It is straightforward to generalize it to a time sequence of operators $\hat{K}_j$.
Then $\sum \hat{K}^\dag_j\hat{K}_j=\hat{1}$ while the probability reads
\begin{equation}
\mathrm{Tr}\,\hat{K}_n\cdots\hat{K}_1\hat{\rho}\hat{K}_1^\dag\cdots\hat{K}_n^\dag
\end{equation}
The choice of the sets $\hat{K}_j$ is quite arbitrary, but a continuous time limit leads to the most natural choice, a Gaussian form
$\hat{K}(a)\propto\exp(-\lambda(a-\hat{A})^2$ for some operator $\hat{A}$ and the outcome $a$  \cite{bfb}. In quantum field theory, this definition can be adapted 
by expressing $\hat{A}$ in terms of local field operators (see Appendix \ref{appa} for the standard quantum field notation conventions). What is even more helpful, Kraus operators can  also be incorporated into the path integral framework
and closed time path (CTP) formalism \cite{schwinger,kaba,matsubara}, see Appendix \ref{appa}. The CTP consists of three parts: the thermal Matsubara part (imaginary) and two flat
 parts (real), forward and backward.
We shall distinguish them by denoting $x_\pm=x\pm i\epsilon$ for an infinitesimally small real positive epsilon. Then the Kraus operators need the field
with the part specified, i.e.
$\hat{K}(x)\to K(x_+)$, $\hat{K}^\dag(x)\to K(x_-)$, see details in Appendix \ref{appa}. In other words, $\hat{K}$, expressed by some field $\phi(x)$ must be placed on the proper part of CTP.
For a moment, the relation between $K$ and $\phi$ is a completely general functional, 
and can be nonlinear and nonlocal, but in short we will heavily reduce this freedom.

The simplest example is the Gaussian form
\begin{equation}
\hat{K}(a)\propto \exp\left[-\lambda\left(\int f(x)\hat{\phi}(x)dx-a\right)^2\right],
\end{equation} ignoring the $\lambda-$dependent global normalization.
We can adapt it to the path integral form, combining both $\hat{K}$ and $\hat{K}^\dag$ into a single form, distinguishing the CTP parts
\begin{equation}
K(a)\propto \exp\left[-\lambda\sum_\pm\left(\int f(x)\phi(x_\pm)dx-a\right)^2\right]
\end{equation}
where $f(x)$ is a real-valued function localized in spacetime (nonzero only inside a finite region of the spacetime) such that $f(x_+)=f(x_-)=f(x)$ 
on the flat part of the CTP. 
The apparent nonlocality of the above construction can be removed by the Fourier transform
\begin{align}
&K(a)\propto\int d\xi_+d\xi_-\;e^{ia(\xi_--\xi_+)-(\xi_+^2+\xi_-^2)/4\lambda}\nonumber\\
&\times\exp\int if(x)(\xi_+\phi(x_+)-\xi_-\phi(x_-))dx 
\end{align}
which has an interpretation of two independent random external fields $\xi_\pm$ for the upper and lower parts of the contour.

We shall apply the measurement to the simple bosonic field with the Lagrangian density
\begin{equation}
2\mathcal L(x)=\partial\phi(x)\cdot\partial\phi(x)-m^2\phi^2(x)\label{bos}
\end{equation}

By Wick theorem \cite{wick}, all correlations can be expressed in terms of products of two-point propagators, $\langle \phi(x)\phi(y)\rangle$.
Simple, translation-invariant correlations read
\begin{equation}
\langle \phi(x)\phi(y)\rangle=Z\int \mathcal D\phi \phi(x)\phi(y)\exp\int i\mathcal L(z)dz
\end{equation}
with special cases defined on the flat part (Im $x\to 0$)
\begin{align}
&S(x,y)=\langle\phi(x_+)\phi(y_+)\rangle=\langle\phi(x_-)\phi(y_-)\rangle^\ast,\nonumber\\
&B(x,y)=\langle \phi(x_+)\phi(y_-)\rangle,\nonumber\\
&C(x,y)=\langle\phi_c(x)\phi_c(y)\rangle,\nonumber\\
&
G(x,y)=\langle \phi(x_\pm)\phi_q(y)\rangle,
\end{align}
for $2\phi_c(x)=\phi(x_+)+\phi(x_-)$ (symmetrized field, classical counterpart)
and $\phi_q(y)=\phi(y_+)-\phi(y_-)$ (antisymmetric, quantum susceptibility to external influence).
Here $S$ is known as Feynman propagator, $C$ is the symmetric correlation (real) while $G$, the causal Green function, is imaginary and
 satisfies $G=0$ for $y^0>x^0$, and $\langle\phi_q(x)\phi_q(y)\rangle=0$. To shorten notation, using the translation invariance (also for complex times),
 we shall identify $X(x,y)\equiv X(x-y)$ for $X=S,B,C,G$.
Note that using the identity $F(x_\pm)=F_c(x)\pm F_q(x)/2$ we get $B(x)=C(x)+(G(-x)-G(x))/2$ and $S(x)=C(x)+(G(x)+G(-x))/2$.
The function $G$ is responsible for causality, i.e. $G(x)\neq 0$ only if $x\cdot x\geq 0$ and $x^0\geq 0$.

To measure the field in a vacuum we can define the probability in terms of path integrals
\begin{equation}
p(a)=Z\int \mathcal D\phi K(a)\exp\int i\mathcal L(z)dz,
\end{equation}
where $Z$ is the path integral normalization, see Appendix \ref{appb}.
The normalization of the probability can be checked by the identity
\begin{equation}
\int da K(a)=\exp\left[-\frac{\lambda}{2}\left(\int f(x)\phi_q(x)dx\right)^2\right]\label{nor}
\end{equation}
The right hand side is $1$
because $\phi(x_+)\equiv \phi(x_-)$ at the \emph{latest} time $x^0$ (they meet at the returning tip of the time path).

It is convenient to introduce the concept of the generating function
\begin{equation}
S(\chi)=\int da p(a)e^{i\chi a}.\label{gen}
\end{equation}
which allows express moments and cumulants
\begin{equation}
\langle a^n\rangle=\left.\frac{d^nS}{d(i\chi)^n}\right|_{\chi=0},\:\langle\langle a^n\rangle\rangle=\left.\frac{d^n\ln S(\chi)}{d(i\chi)^n}\right|_{\chi=0}.
\label{momcum}
\end{equation}
They are related, in particular,
\begin{align}
&\langle\langle a\rangle\rangle=\langle a\rangle,\;\langle\langle a^2\rangle\rangle=\langle\delta a^2\rangle,
\nonumber\\
&\langle\langle a^3\rangle\rangle=\langle \delta a^3\rangle,\;\langle\langle a^4\rangle\rangle=\langle\delta a^4\rangle-3\langle\langle a^2\rangle\rangle^2
\end{align}
for $\delta a=a-\langle a\rangle$. Gaussian distributions have only nonzero first two cumulants.
It is helpful to introduce convolutionwith a \emph{bare} quasiprobability $p_\bullet$ stripped from the pure Gaussian noise depending on the 
strength of the measurement $\lambda$, namely
\begin{equation}
p(a)=(2\lambda/\pi)^{1/2}\int d\bar a\, e^{-2\lambda(a-\bar{a})^2}p_\bullet(\bar a)
\end{equation}
which leads to
$
\ln S(\chi)=\ln S_\bullet(\chi)-\chi^2/8\lambda
$
so only $\langle\langle a^n\rangle\rangle=\langle\langle a^n\rangle\rangle_\bullet$ for $n\neq 2$ while
$
\langle\langle a^2\rangle\rangle=\langle\langle a^2\rangle\rangle_\bullet+1/4\lambda
$. We shall mark with $\bullet$ all averages and generating functions involving $p_\bullet$ to
distinguish it from $p$. In other words, we can separate the measurement statistics into the Gaussian detection noise of the variance $1/4\lambda$,
divergent in the limit $\lambda\to 0$, and the bare function that turns out to be a quasiprobability. It is normalized
and has well defined moments but lacks general positivity.

In the case of our measurement, we have a Gaussian function
\begin{align}
&S_\bullet(\chi)=Z\int \mathcal D\phi\exp\int i\mathcal L(z)dz\nonumber\\
&\times\exp\left[-\frac{\lambda}{2}\left(\int f(x)\phi_q(x)\right)^2\right]
\exp \int i\chi f(y)\phi_c(y)dy.
\end{align}
Therefore $\langle a\rangle=0$ in the vacuum and 
\begin{align}
&\langle a^2\rangle_\bullet=Z\int \mathcal D\phi \exp\int i\mathcal L(x)dx\nonumber\\
&\times \exp\left[-\frac{\lambda}{2}\left(\int f(x)\phi_q(x)dx\right)^2\right]\left(\int f(y)\phi_c(y)dy\right)^2
\end{align}
Now, the term $\phi(x_+)-\phi(x_-)$ must be contracted with some $\phi(y)$ with $y^0>x^0$; otherwise, it vanishes. It leaves essentially only a few terms
\begin{align}
&\langle a^2\rangle_\bullet=\int dxdy f(x)f(y)C(x-y)\nonumber\\
&-\lambda\left(\int dxdy f(x)f(y)G(x-y)\right)^2.
\end{align}

Suppose now that we perturbed the vacuum. This is the natural physical situation when
a beam of particles is sent to the detector. Let
\begin{equation}
\exp\int idz\mathcal L(z)\to\mathcal P[\phi]\exp\int idz\mathcal L(z) \label{per1}
\end{equation}
where $\mathcal P$ denotes perturbation
\begin{equation}
\mathcal P[\phi]=\exp\int  ig(w)\phi_q(w)dw\label{per2}
\end{equation} by the shift of the field induced by $g(w)$ (localized in spacetime).
In principle, we should discuss not only the particle detector but also the generator.
Since we prefer to focus on the detection part we stay at the minimal description with the single perturbation function $g$.
Almost all the above formulas remain valid, i.e. probability has the same $g$-independent normalization and is Gaussian. It only
gets the nonzero average. We calculate
\begin{align}
&S_\bullet(\chi)=Z\int \mathcal D\phi\exp\int i\mathcal L(z)dz\times\nonumber\\
&\exp\int ig(w)\phi_q(w)dw\times\nonumber\\
&\exp\left[-\frac{\lambda}{2}\left(\int f(x)\phi_q(x)dx\right)^2\right]
\exp \int i\chi f(y)\phi_c(y)dy.
\end{align}
The average is $\lambda$-independent while the higher cumulants remain unaffected,
\begin{equation}
\langle a\rangle_\bullet=\int dwdy f(y)g(w)iG(y-w).\label{fgg}
\end{equation}
As we see, the linear measurement gives simply Gaussian statistics and cannot be used to model single particle detection
with the Poissonian clicks.
Nevertheless, already the self-consistency of the above construction is a promising signature that the approach by Kraus operators is correct.

\subsection{Comparison to Unruh-DeWitt model}
The original Unruh-DeWitt model
\cite{unruh,witt} can be compared with the Kraus one by taking $f(x)=\int F(\tau)\delta(x-y(\tau))d\tau$
for a certain world line $y$ along the proper time $\tau$, where $F$ is not a function but another field with its own Lagrangian part. Already in the Kraus model a trivial $y(\tau)=(\tau,0,0,0)$ 
would lead to divergence of (\ref{fgg}). It can be cured by regularization,
\cite{taka,schlicht,langlois,louko}, or replacing $f$ by another auxiliary field \cite{gale}.
In all these approaches one constructs the indirect measurement involving only $F$ and not the original $\phi$.
Therefore, our approach is a shortcut. Instead of any auxiliary objects, one defines the Kraus functional,
depending not on fields but on usual functions and spacetime. In principle, one can map each type of Unruh-DeWitt model
onto some Kraus form but we claim that starting directly for Kraus is simpler and makes further calculation more manageable.

\section{Nonlinear  measurement}

We shall define a quadratic Kraus operator and apply it to the vacuum and continuous plane wave, showing that it 
exhibits features of Poisson statistics in contrast to the Gaussian linear case.

\subsection{Quadratic measurement}

Let us define Kraus operators in terms of path integrals
\begin{equation}
K(a)\propto\exp\left[-\lambda\sum_\pm\left(\int f(x)\phi^2(x_\pm)dx-a\right)^2\right],
\end{equation}
which can be made local as previously by a Fourier transform.
Analogously to the field measurement, we can write down the formal expression for the generating function, namely
\begin{align}
&S_\bullet(\chi)=Z\int \mathcal D\phi\exp\int i\mathcal L(z)dz\nonumber\\
&\times\exp\left[-\frac{\lambda}{2}\left(\int f(x)\phi^2_q(x)dx\right)^2\right]
\exp \int i\chi f(y)\phi^2_c(y)dy
\end{align}
where $\phi^2_c=(\phi_c)^2+(\phi_q)^2/4$ and $\phi^2_q=2\phi_c\phi_q$.

Our aim is to calculate $\langle a^n\rangle$ and show that for some linear perturbation $g$, 
there exists a regime (some $f$) where
the distribution is Poissonian,
\begin{equation}
p(a=n\eta)=e^{-\alpha}\alpha^n/n!,
\end{equation}
proving approximate dichotomy $a=0,\eta$ from $\langle a^2(a-\eta)^2\rangle\simeq 0$. 
Poisson distribution has simple cumulants (\ref{momcum})
$\langle\langle a^n\rangle\rangle=\alpha\eta^n$
but  they coincide with the moments for $\alpha\ll 1$.
We will attempt to expand $\langle a^n\rangle$ in  powers $\lambda$ and estimate the upper bound for the higher terms of the expansion using the Wick theorem.

\subsection{The vacuum}

We shall apply the quadratic measurement to the vacuum.
It is then useful to calculate moments,
\begin{align}
&\langle a^n\rangle_\bullet=Z\int \mathcal D\phi\exp\int i\mathcal L(z)dz\nonumber\\
&\times\exp\left[-\frac{\lambda}{2}\left(\int f(x)\phi^2_q(x)dx\right)^2\right]
\left(\int f(y)\phi^2_c(y)dy\right)^n\label{coa}
\end{align}
expanding in powers of $\lambda$. 
Problems arise when $\phi^2$ contains two fields at the same time so, e.g., $\langle \phi^2(x)\rangle=\langle \phi^2_c(x)\rangle\to \infty$
It must be \emph{renormalized}, e.g., by subtracting the counteraverage  for a fictitious mass $M\to\infty$.
We can subtract large masses i.e.
\begin{equation}
\langle \phi^2_c(x)\rangle\to  \langle \phi^2_c(x)\rangle+\sum _j \epsilon_j\langle \phi^2_c(x)\rangle_{m\to M_j}=\Lambda
\end{equation}
where $M_j\gg m$ are large renormalization masses (Pauli-Villars) \cite{pavi,peskin} while $\epsilon_j$ are some numbers (e.g. $\pm 1$) not too large.
The constant $\Lambda$ is an unobservable calibration shift.
From now on we also make this shift in $a$, i.e. 
$
a\to a-\Lambda\int f(x)dx
$.

In the lowest order of $\lambda$
\begin{equation}
\langle a^2\rangle_\bullet=\int f(x)f(y)\langle \phi^2_c(x)\phi^2_c(y)\rangle dxdy
\end{equation}
Denoting
$
\tilde f(k)=\int dx e^{ik\cdot x}f(x)/(2\pi)^{D+1}
$
with the help of
\begin{align}
&\langle \phi^2_c(x)\phi^2_c(y)\rangle=\langle \phi^2_c(x)\phi^2_c(y)\rangle-\langle \phi^2_q(x)\phi^2_q(y)\rangle/4
\nonumber\\
&=\langle\phi^2(x_+)\phi^2(y_-)+\phi^2(x_-)\phi^2(y_+)\rangle/2
\end{align}
  we get (see   Appendix \ref{appc}),
\begin{equation}
\langle a^2\rangle_\bullet=
\int W(q)|\tilde f(q)|^2dq \label{aa0}
\end{equation}
with 
\begin{equation}
W(q)=\frac{\pi^2(q\cdot q/4-m^2)^{(D-2)/2}S_D}{\sqrt{q\cdot q}}\label{www}
\end{equation}
for $q\cdot q>4m^2$ and $0$ otherwise. Here $S_D=2\pi^{D/2}/\Gamma(D/2)$ is the surface of a unit ball in $D$ dimensions.
 It shows that nonlinear fluctuations of the field need pair creation. For $f$ varying over
time/length much larger than $1/m$, they are negligible, allowing low-noise measurement in the vacuum.

\subsection{Measurement of the plane wave}

We want the perturbation $g$ to generate an enveloped wave of a particular frequency, i.e. of the form
\begin{equation}
g(x)=e^{iE_px^0}h(x^1+L)+e^{-iE_px^0}h^\ast(x^1+L)
\end{equation}
with a function $h(y)$ localized  at $|y|\ll L$ and $E_p>m$,
which should generate a plane wave in the $x^1$ direction.
It corresponds to a constant coherent flux of free particles in the $x^1$ direction.
Our measurement model is able to capture single particles in the flux in contrast to the vacuum and its fluctuations.
The effect of the perturbation can be described by
\begin{equation}
G_g(x)=\int dy G(x-y)g(y)
\end{equation}
in the limit $L\to\infty$, where it reads (see Appendix \ref{appd})
\begin{equation}
G_g(x)=2i\mathrm{Im}Ae^{i|p|x^1-iE_px^0}
\end{equation}
or equivalently
\begin{align}
&G_g(x)=G_{g+}(x)+G_{g-}(x),\nonumber\\
&
G_{g\mp}=(iA_i\pm A_r)e^{\mp ip\cdot x},\label{gga}
\end{align}
for $p=(E_p,|p|,0,0)$ and $2A=\tilde{h}(-|p|)/|p|$. 

For  a linear perturbation $g$ we can now determine the measurement statistics, inserting (\ref{per2})
into (\ref{coa}).

The measurement function $f(x)$ will vary at the scale much longer than $1/E_p$.
Defining 
\begin{equation}
F(x)=\sum_{\mp} e^{\mp ip\cdot x}F_\mp(x),
\end{equation}
for $F=G,C,S,B$, and expanding $k=k'\mp p$ in (\ref{zzz}) for small $k'$
\begin{align}
&C_\mp(x)=\int \frac{ dk'}{2(2\pi)^{D}}\delta(2k'\cdot p)e^{ik'\cdot x},\nonumber\\
&G_\mp(x)=\int\frac{idk'}{(2\pi)^{D+1}}\frac{e^{ik'\cdot x}}{\mp 2p\cdot k'_+},
\end{align}
explicit calculation gives
\begin{align}
&B_+(x)=\delta(x^\perp)\delta(x^0v-x^1)/2E_p,
\nonumber\\
&C_\mp(x)=B_+(x)/2,\nonumber\\
&G_\mp(x)=\pm \theta(x^0)B_+(x),\nonumber\\
&S_\mp(x)=\theta(\pm x^0)B_+(x),\label{axx}
\end{align}
and $B_-=0$
with $x^\perp=(x^2,x^3,\dots)$ and
 the speed of the field  $v=|p|/E_p$ (in the units of the speed of light).
 This is a very intuitive physical picture since the dynamics is concentrated on the lines of the propagation at constant speed $v$.

\begin{figure}
\includegraphics[scale=.4]{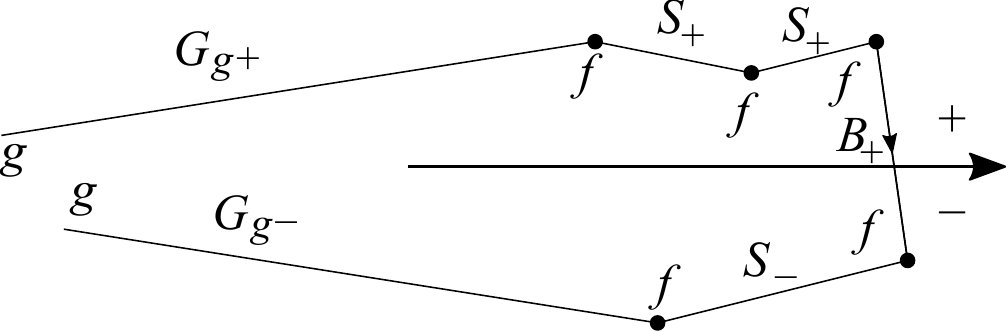}
\caption{A chain of propagators $G$, $S$ and $B$. The arrow points into the time direction while positive and negative imaginary sides are denoted by $+$ and $-$, respectively.}
\label{gsbf}
\end{figure}

In the lowest order of $g$, the average
$
\langle (g\phi)(g\phi)(f\phi^2)\cdots (f\phi^2) \rangle
$
turns out to be a sum of Feynman graphs with part of the vertices on the $+$ side and part on the $-$ side of CTP,
see Fig. \ref{gsbf}.

We shall now express the first $\langle a^n\rangle_0$, $n=1,2,3,4$ in terms of the just derived functions
in the lowest order of $\lambda$, and we get
\begin{align}
&\langle a^n\rangle_\bullet\simeq 2E_p|A|^2\nonumber\\
&\times\int d\boldsymbol x\left(\int dx^0 f(x^0,x^1+vx^0,x^\perp)/E_p\right)^n,\label{fan}
\end{align}
where  we used the shift $x^1\to  x^1-vx^0$ in the integrals.
The factor $2E_p$ appears because each $\phi^2$ contains $2$ fields $\phi$ giving $2$ per vertex and $2^n$ in total.
On the other hand $2\phi^2=\phi^2(x_+)+\phi^2(x_-)$ so we get a factor $2^{-n}$ to cancel out with the above one.
Each line ($S_\pm$ and $B_+$) has the factor $(2E_p)^{-1}$ giving $(2E_p)^{1-n}$ in total. For $n$ vertices we have all possible 
decompositions into $n_+$ and $n_-=n-n_+$ vertices with $n_+=0\cdots n$. All combinations give the total factor $(1+1)^n=2^n$
(by the binomial formula). There are no additional factors $n_{\pm}!$ because these factors cancel out
($n_\pm$ permutations cancel out with $\theta(\pm x^0)$ ordering). 

\begin{figure}
\includegraphics[scale=.5]{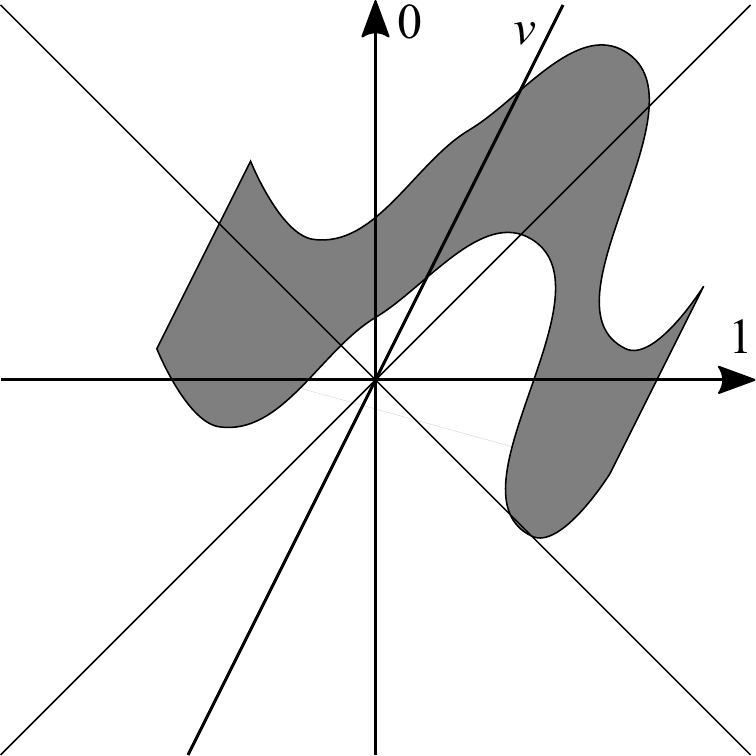}
\caption{Projecting of the measurement lines onto the $D$-dimensional spatial base}
\label{proj}
\end{figure}

\begin{figure}
\includegraphics[scale=.5]{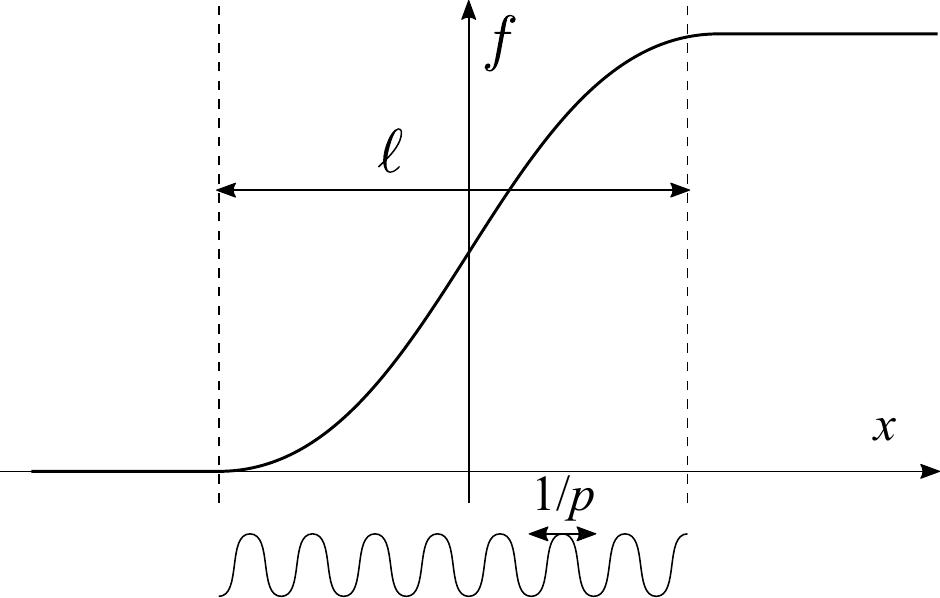}
\caption{Length scales in measurement. The wavelength $1/p$ is shorter than the variation length $\ell$ of the envelope function $f$.}
\label{dxp}
\end{figure}

To get the Poisson statistics, it is sufficient that
\begin{equation}
\int dx^0 f(x^0,x^1+vx^0,x^\perp)/E_p=\left\{\begin{array}{ll}
\eta&\mbox{ if }\boldsymbol x\in V\\
0&\mbox{ if }\boldsymbol x\notin V\end{array}\right.
\end{equation}
where $V$ is a certain volume in $D$ dimensional space.
In other words, we need a constant integral of the measuring function $f$ along the line of speed $v$, i.e. 
$x^1=vx^0$ (see Fig. \ref{proj}). We have also $\alpha=2E_p|A|^2V$.
The approximation is valid as long as the variation length scale of $f$, say $\ell$ is much larger than
the wavelength $1/p$, i.e., $p\ell \gg 1$ (see Fig. \ref{dxp}).

\begin{figure}
\includegraphics[scale=.4]{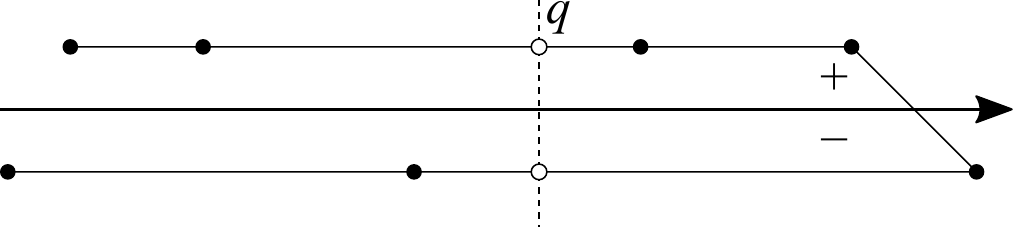}
\caption{Approximate cancellation of disturbances caused by the measurement. The insertion of $\phi^2_q$  at $x_+$ and $x_-$ in the chains of propagators (\ref{axx})
on the line of propagation gives two exactly opposite terms.}
\label{gqq}
\end{figure}

Finally, higher order terms in the $\lambda$ expansion will contain $\phi^2_q=\phi^2(x_+)-\phi^2(x_-)$. 
Fortunately, in our approximation  (\ref{axx}) these terms cancel. This is because inserting such points in the existing graph gives 
always two opposite expressions (see Fig. \ref{gqq}).

\subsection{
Perturbative estimates}

\begin{figure}
\includegraphics[scale=.5]{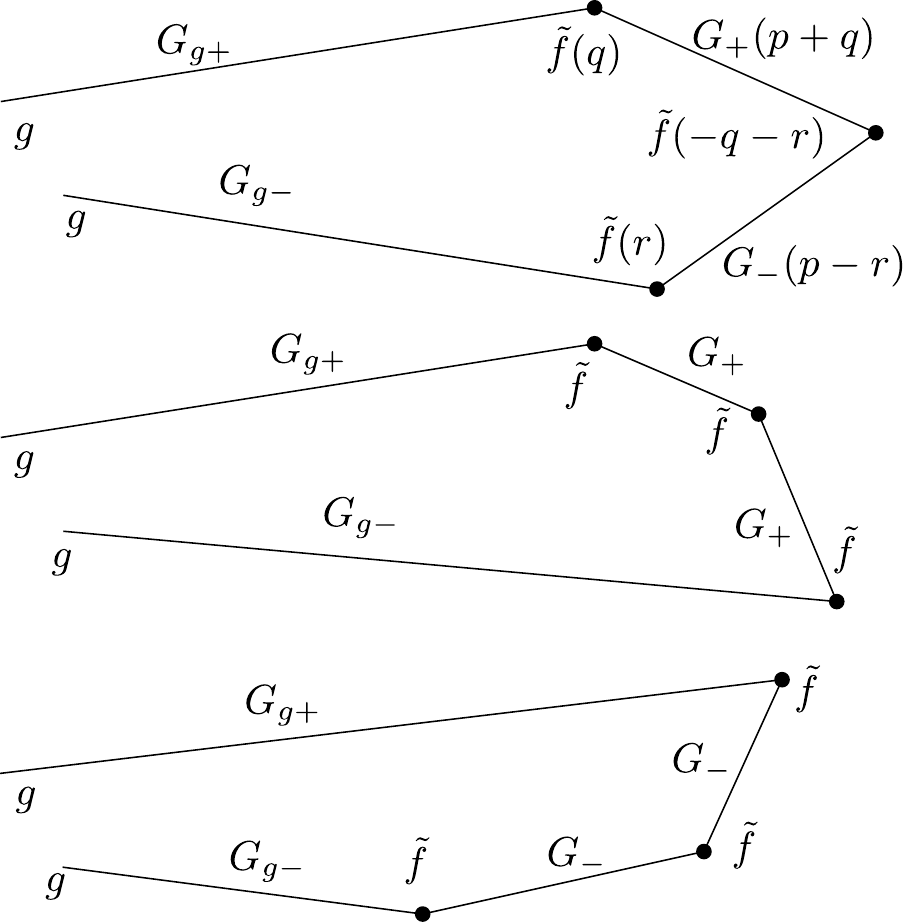}
\caption{The Feynman-Schwinger graph contributing to the $\lambda$ correction to $\langle a\rangle$}
\label{gg2}
\end{figure}

We shall calculate $\langle a\rangle$ including order $\lambda$.
In the leading order
$$
\langle a\rangle=2|A|^2\int dx f(x^0,x^1+vx^0,x^\perp)=2|A|^2\int dx f(x)
$$
because the shift is irrelevant. This result is exact in the leading order, $\lambda^0$. 
Now, the next order term, $\lambda$ would be $0$ if we make our approximations on $G$, $S$, $B$.
To find a nonzero contribution, we need to estimate the lowest deviations.
The Feynman-Schwinger graph is depicted in Fig. \ref{gg2}.

The deviation of $G$  reads
\begin{equation}
\Delta G_\mp(x)=\int\frac{dk'}{(2\pi)^{D+1}}\frac{k'\cdot k' e^{ik'\cdot x}}{i(2p\cdot k'_+)^2}
\end{equation}
for $k^{\prime 0}=k^{\prime 1}|p|/E_p$ and $\partial=\partial_x$.
Hence,
\begin{equation}
\Delta G_\mp(x)=\partial\cdot\partial[\theta(x^0)x^0B_+(x)]/2iE_p
\end{equation}
The derivatives can be moved to $f$ so the graphs will contain
products $(\partial f)(\partial ff)$ or $(\partial\partial f)ff$.

We shall estimate the corrections for 
\begin{equation}
f(x)=F I_0(x^0-x^1/v-w(\boldsymbol x))I(\boldsymbol x)]\label{fan2}
\end{equation}
where $I_0(t)=\theta(t)\theta(L_0-t)$ and
\begin{equation}
I(\boldsymbol x)=\left\{\begin{array}{ll}
1&\mbox{ if }\boldsymbol x\in V\\
0&\mbox{ if }\boldsymbol x\notin V\end{array}
\right.
\end{equation}
 of linear size $L$ for some function (measurement start time) $w$ defined inside $V$. In this case, the derivatives are nonzero only on the boundary.
In the case of two first derivatives, it gives essentially a product of two $\delta$ functions, pinning $f$ to the boundary.
It turns out that the case of the second derivative is actually of the same order. 
Formally we would get $\delta'$ which is infinite, so in both cases we need to regularize $\theta$ functions. 
Let us assume that $\theta$ and $I$ change from $0$ to $1$ smoothly over the distance $\ell\ll L,L_0$. 
The total correction to $\langle a\rangle$ is
$
\sim\lambda\langle a^3\rangle_0 L_0/L\ell E_p
$
since the numerator contains also $x^0\sim L_0$. 
Therefore, the Poisson statistics will be a good approximation in the limit
$
\lambda\eta^2\ll E_p\ell L/L_0
$
with $\eta=F L_0/E_p$. On the other hand, the detection noise cannot blur the distribution giving another constraint
$
1\ll \lambda (\alpha\eta)^2
$.
If we want maximally a single click, then additionally
$
\alpha\ll 1
$.
Summarizing, the single click statistics occurs if
\begin{equation}
1\ll\alpha^{-2}\ll\lambda\eta^2\ll\ell LE_p/L_0\label{comp}
\end{equation}

Finally, the previously calculated vacuum fluctuation should also be small. They are almost completely negligible (exponentially)
if $m\ell\gg 1$, i.e. the measurement shape function varies over scales impossible to generate pairs of particles. However, in the massless
 (or low mass) case, the fluctuations are estimated by $W\sim (q\cdot q)^{(D-3)/2}$ giving $|q|\sim 1/\ell$, i.e.
$
\langle a^2\rangle_0\sim \ell^{2-2D}\tilde f^2(0)
$
 and $\tilde f(0)\sim VFL_0$. At $D=1$ they are actually divergent logarithmically with the shrinking mass $m$. 
For $D>1$ we need also $VE_p/\ell^{D-1}\ll\alpha$. Together with the previous condition it implies
$
\ell LE_p/L_0\gg 1\gg VE_p/\ell^{D-1}
$
which gives the impossible requirement $\ell^D \gg VL_0/L$. The reason for this failure is that $\phi^2$ is not a conserved quantity and the measurement
can easily change it locally (above the mass threshold). We shall resolve this problem by replacing $\phi^2$ by the conserved energy-momentum density $T_{\mu\nu}$.

\section{
Energy-momentum measurement}

We shall modify the previous simple quadratic measurement replacing $\phi^2$ by energy-momentum density, which is still quadratic
in $\phi$ but the additional derivatives turn out to suppress unwanted noise in the high energy limit.

\subsection{Energy-momentum tensor}
\label{emt}

\begin{figure}
\includegraphics[scale=.4]{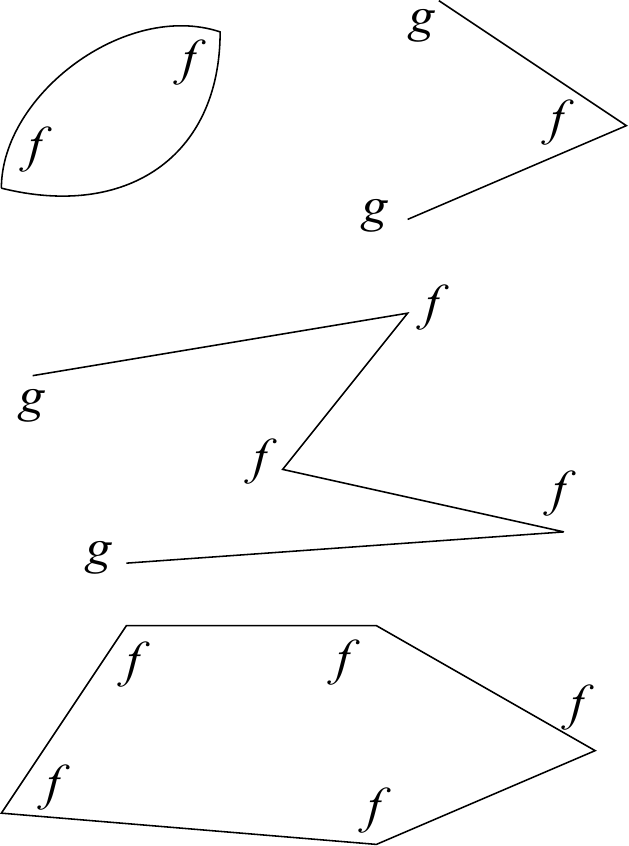}
\caption{Loops and chains in the graphs involving energy-momentum tensor. They vanish if solely $T_q$ (not $T_c$) appears in a loop/chain}
\label{split}
\end{figure}

Energy-momentum stress tensor by Noether theorem reads
\begin{equation}
T^{\mu\nu}=\frac{\partial \mathcal L}{\partial(\partial_\mu\phi)}\partial^\nu\phi-g^{\mu\nu}\mathcal L
\end{equation}
with $\partial^\nu=g^{\nu\tau}\partial_\tau$
equal in our case
\begin{equation}
T^{\mu\nu}=\partial^\mu\phi\partial^\nu\phi-g^{\mu\nu}(g^{\sigma\tau}\partial_\sigma\phi\partial_\tau\phi-m^2\phi^2)/2.
\end{equation}
We define the energy-momentum measurement
\begin{equation}
K(a)\propto\exp\left[-\lambda\sum_\pm\left(\int f_{\mu\nu}(x)T^{\mu\nu}(x_\pm)dx-a\right)^2\right],
\end{equation}
which is normalized in the same way as the field, with symmetric $f_{\mu\nu}=f_{\nu\mu}$.
The generating function reads
\begin{align}
&S_\bullet(\chi)=Z\int \mathcal D\phi\exp\int i\mathcal L(z)dz\nonumber\\
&\times\exp\left[-\frac{\lambda}{2}\left(\int f_{\mu\nu}(x)T^{\mu\nu}_q(x)dx\right)^2\right]\nonumber\\
&\times
\exp \int i\chi f_{\mu\nu}(y)T^{\mu\nu}_c(y)dy,
\end{align}
where we denoted $T^{\mu\nu}_q(x)=T^{\mu\nu}(x_+)-T^{\mu\nu}(x_-)$
and $2T^{\mu\nu}_c(x)=T^{\mu\nu}(x_+)+T^{\mu\nu}(x_-)$.
Note also that
\begin{align}
&2T^{\mu\nu}_q(x)=\partial^\mu\phi_c\partial^\nu\phi_q+\partial^\mu\phi_q\partial^\nu\phi_c\nonumber\\
&-g^{\mu\nu}(\partial\phi_c\cdot\partial\phi_q-m^2\phi_c\phi_q),\nonumber\\
&T^{\mu\nu}_c(x)=\partial^\mu\phi_c\partial^\nu\phi_c-g^{\mu\nu}(\partial\phi_c\cdot\partial\phi_c-m^2\phi_c^2)/2\nonumber\\
&+\partial^\mu\phi_q\partial^\nu\phi_q/4-g^{\mu\nu}(\partial\phi_q\cdot\partial\phi_q-m^2\phi_q^2)/8.
\end{align}
As in the case of $\phi^2$, the calculations involve correlations of the type
$
\langle (g\phi)\cdots (g\phi) (fT)\cdots(fT)\rangle
$.
However, there are dangerous contact terms to be regularized by fermionic ghosts \cite{peskin,fadeev}, see Appendix \ref{appe}.
Fortunately, once identified, we can basically forget about ghosts, and just keep the unitarity constraints when calculating loops, 
$\langle T_q(x)T_q(y)\cdots T_q(w)\rangle=0$
as a calculation rule if only $T_q$ are involved. Basic examples of graphs involved in our calculations are depicted in Fig. \ref{split}.
From now on we shall subtract the zero-temperature average from $T$, i.e. $T\to T-\langle T\rangle_0$,
as it is unobservable and contains the renormalization parameters, and we are interested only in
the noise and sensitivity of the detector to the incoming particles. By this shift $\langle T\rangle=0$.

\subsection{Measurement of the vacuum}

In the lowest order of $\lambda$
\begin{equation}
\langle a^2\rangle_\bullet=\int f_{\mu\nu}(x)f_{\xi\eta}(y)\langle T^{\mu\nu}_c(x)T^{\xi\eta}_c(y)\rangle dxdy
\end{equation}
analogously to the $\phi^2$. We get (see Appendix \ref{appf})
\begin{equation}
\langle a^2\rangle_\bullet=\int dq\,\tilde f_{\mu\nu}(q)\tilde f_{\xi\eta}(-q)\frac{X^{\mu\nu\xi\eta}(q)W(q)}
{(q\cdot q)^2(D+2)D}\label{ctt}
\end{equation}
with \cite{vac}
\begin{align}
&X^{\mu\nu\xi\eta}=
P(q^\mu q^\nu -(q\cdot q) g^{\mu\nu})(q^\xi q^\eta-(q\cdot q)g^{\xi\eta})
\nonumber\\
&+R[(q^\mu q^\eta -(q\cdot q) g^{\mu\eta})(q^\xi q^\nu-(q\cdot q)g^{\xi\nu})\nonumber\\
&+
(q^\mu q^\xi -(q\cdot q) g^{\mu\xi})(q^\nu q^\eta-(q\cdot q)g^{\nu\eta})],\nonumber\\
&P=m^4+(D+1)m^2q\cdot q/2+(q\cdot q)^2(D^2-3)/16,
\nonumber\\
&R=m^4-m^2q\cdot q/2+(q\cdot q)^2/16,\label{ttt}
\end{align}
and  $W$ defined by (\ref{www})
The case $D=1$ is degenerate, see Appendix \ref{appf}.

\subsection{Poisson statistics}

We can adapt most of the results from the $\phi^2$ replacing $f$ with $f_{\mu\nu}$ and adding factor $p^\mu p^\nu$ from $T$ in (\ref{fan}).
Then e.g. $\eta=\int dx^0 f_{\mu\nu}(x^0,x^1+vx^0,x^\perp)p^\mu p^\nu/E_p$. Replacing further $F$ with $F_{\mu\nu}$ in (\ref{fan2})
we have $\eta=F_{\mu\nu}p^\mu p^\nu L_0/E_p$. We only need to include deviation from derivatives in $T$ that can act on variables in $f$
considering potential disturbance (the $\sim\lambda$ term).
Fortunately, if, say, $F_{00}=F$ and $0$ otherwise (assuming we measure energy density, not momentum), their contribution to the disturbance is negligible,
and we stay with (\ref{comp}).

What is qualitatively different,  vacuum fluctuations remain small in the massless limit.
From our above discussion we have then $\langle a^2\rangle\sim \ell^{-2(D+1)}\tilde F^2(0)$.
If  we want
$
\ell^{-D-1}\tilde F\ll \alpha\eta
$ then $V/\ell^{D+1}E_p\ll \alpha$. In contrast to the $\phi^2$ case increasing $E_p$ helps to satisfy this and the other requirements,
due to the fact that we work with the conserved quantity.

\section{Conclusion}
We have presented a self-consistent measurement model suitable for high energy particle detectors.
Defined completely within the existing framework of quantum field theory combined with POVM and Kraus functionals,
it allows one to identify a particle as a click with almost perfect efficiency, in contrast to the Unruh-DeWitt model.
The presented examples stress the importance of nonlinearity and
connection with conservation principles to choose a proper Kraus functional. 
The strength of the measurement, parameter $\lambda$, must be neither too low, to keep the detection noise small, nor too high, to
keep the backaction small. Only an intermediate regime, depending on the energy, time, and length scales,
 allows $\sim 100\%$ efficiency of the detection.
Although we analyzed a simple bosonic field,
the ideas are quite general and can easily be extended to fermions and compound particles. 
The actual form of the correct Kraus functional can be inferred for the actual experiment by calibration. Assuming some restricted family of parameters, they can be determined in diagnostic tests and later used in other experiments. In more complicated gauge theories, one can still use Kraus functionals in a perturbative fashion
as their positivity is just a formal expectation, to be restored by e.g. families of renormalizing ghost fields as we did in Sec. \ref{emt} and Appendix \ref{appe}. We believe that
further work on such models will help to establish a Bell-type family of experiments in the high energy regime,
identifying the main technical challenges. One can also explore completely different detector's functions $f$, e.g.
 an analog of an accelerating observer as in the original Unruh model or a sequence of measurements. 
Note also that our model is a theoretical idealization and it may need practical adjustments
taking into account specific experiments.

\section*{ACKNOWLEDGEMENTS}
I thank W. Belzig and P. Chankowski for many fruitful discussions on the subject and pointing out important issues.

\appendix
\section{CLOSED TIME PATH FORMALISM}
\label{appa}

We shall summarize the notation in
the quantum field theory of a scalar field, in $D+1$ dimensions ( $D$ spatial dimensions and $1$ time, with $D=3$ in full space but also $D=1$ for simple illustrative cases)
$x=(x^0=ct,\boldsymbol x)$, with time $t$, speed of light $c$ and spatial position $\boldsymbol x=(x^1,\dots ,x^D)$. For simplicity $c=\hbar=1$.
We denote partial derivatives $\partial_\mu=\partial/\partial x^\mu$ and Minkowski scalar product $A\cdot B=A^\mu B_\mu=A^{\mu}g_{\mu\nu}B^\nu$
with flat metric $g^{\mu\nu}=g_{\mu\nu}=1$ for $\mu=\nu=0$, $g^{\mu\nu}=g_{\mu\nu}=-1$ for $\mu=\nu=1\cdots D$ and $g^{\mu\nu}=g_{\mu\nu}=0$ for $\mu\neq\nu$.
Real scalar field $\hat{\phi}(\boldsymbol x)$ with conjugate field $\hat{\pi}(\boldsymbol x)$ obeys commutation relation
\begin{equation}
[\hat{\phi}(\boldsymbol x),\hat{\pi}(\boldsymbol y)]=i\delta(\boldsymbol x-\boldsymbol y)\label{com}
\end{equation}
for $[\hat{A},\hat{B}]=\hat{A}\hat{B}-\hat{B}\hat{A}$.
Tbe relativistic field Hamiltonian reads
\begin{equation}
\hat{H}=\int d\boldsymbol x(\hat{\pi}^2(\boldsymbol x)+|\nabla \hat{\phi}(x)|^2+m^2\hat{\phi}^2(\boldsymbol x))/2.
\end{equation}
Here the $\nabla$ term is, in fact, a sum of partial derivatives
\begin{equation}
|\nabla \hat{\phi(x)}|^2=\sum_{j=1}^D(\partial_j\hat{\phi}(\boldsymbol x))^2.
\end{equation}
The Heisenberg picture transforms the field with time
\begin{equation}
\hat{\phi}(x)=e^{i\hat{H}t}\hat{\phi}(\boldsymbol x)e^{-i\hat{H}t}
\end{equation}
Translation into path integrals gives
\begin{equation}
\langle\Phi'|\exp(-i\hat{H}t)|\Phi\rangle=\int \mathcal D\phi \exp\int i\mathcal L(x)dx 
\end{equation}
with $\phi(x^0=0,\dots)=\Phi$ and $\phi(x^0=t,\dots)=\Phi'$,
where the Lagrangian density $\mathcal L$ is given by (\ref{bos}).

For the fermionic fields, here used only to generate renormalization  counterterms, one has to replace commutator $[\hat{\phi},\hat{\pi}]$ in (\ref{com}) by
anticommutator $\{\hat{\phi},\hat{\pi}\}=\hat{\phi}\hat{\pi}+\hat{\pi}\hat{\phi}$ and introduce Grassmann anticommuting fields $\phi$ in path integrals,
i.e. $\phi(x)\phi(y)=-\phi(y)\phi(y)$, $\int d\phi=0$, $\int \phi d\phi=1$.
The complications, including the signs and order conventions, are thoroughly described in the literature \cite{peskin}.

\begin{figure}
\includegraphics[scale=.5]{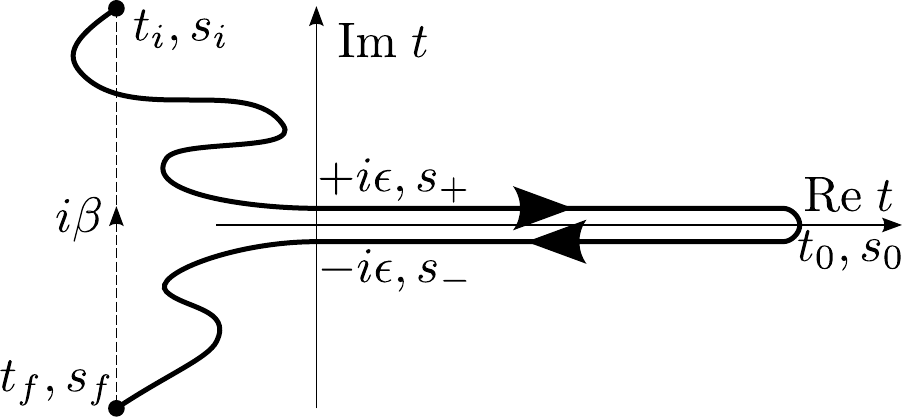}
\includegraphics[scale=.5]{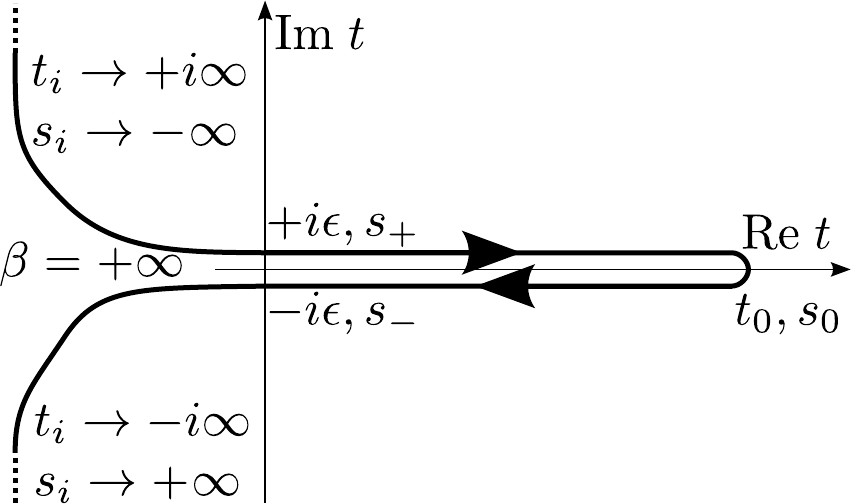}
\caption{The time path in the CTP approach in the case of finite temperature $\beta=1/T$. At zero temperature, the shift $\beta$ stretches to infinity
with $t_i\to +i\infty$, $t_f\to -i\infty$. }\label{ctp}
\end{figure}

Time flow over the Schwinger-Kadanoff-Baym-Matsubara CTP \cite{schwinger, kaba,matsubara,weldon,chou,landsman,kapusta}
is parametrized $t(s)$ (optionally 
with a subscript indicating the specific point in spacetime, $x^0(s_x)$). The real parameter $s\in [s_i,s_f]$ must satisfy $dt/ds\neq 0$ 
and $\mathrm{Im}\;dt/ds\leq 0$, the jump $t(s_i)-t(s_f)=i\beta$ for $\beta=1/k_BT>0$ (inverse temperature); see Fig. \ref{ctp}.
For fermionic fields, the jump is accompanied by the sign reversal for each field.
In the case $k_BT\to 0$ we have $t(s_\mp)\to \pm i\infty$. For convenience the flat part splits into $t\to t_{\pm}=t(s_\pm)=t\pm i\epsilon$ ($\epsilon\to 0_+$, a small positive number going to $0$ in the limit) and $x_\pm=(t_\pm,\boldsymbol x)$ with $s_+<s_-$.

We use the derivative rule: 
$
\partial_0=(dt/ds)^{-1}\partial/\partial s
$
and differential $dx=dx^0dx^1\cdots dx^D$ with $dx^0=(dt/ds)ds$ and
\begin{equation}
\delta(x-y)=\delta(x^0-y^0)\delta(x^1-y^1)\cdots\delta(x^D-y^D)
\end{equation}
with $\delta(x^0-y^0)=\delta(s_x-s_y)/(dt/ds)|_{s=s_x=s_y}$. These rules allow one to establish the full compliance  of CTP with perturbative relativistic quantum
field theory \cite{ab13,ab16}.

\section{TWO-POINT CORRELATIONS}
\label{appb}

With the definitions in Appendix \ref{appa} and Eq. (\ref{bos}) one can calculate all relevant quantum field theory functions, i.e.
\begin{align}
&\langle \phi(x)\phi(y)\cdots\rangle=\langle \mathcal T\hat{\phi}(x)\hat{\phi}(y)\cdots\rangle\nonumber\\
&=Z\int \mathcal D\phi \phi(x)\phi(y)\exp\int i\mathcal L(z)dz
\end{align}
with
\begin{equation}
Z^{-1}=\int \mathcal D\phi \exp\int i\mathcal L(z)dz,
\end{equation}
where $\mathcal T$ denotes ordering by $s$, i.e.
\begin{equation}
\mathcal T\hat{\phi}(x)\hat{\phi}(y)=\left\{\begin{array}{ll}
\hat{\phi}(x)\hat{\phi}(y)&\mbox{ if }s_x>s_y,\\
\hat{\phi}(y)\hat{\phi}(x)&\mbox{ if }s_y>s_z.\end{array}\right.
\end{equation}

Since the  formal path functional is Gaussian, all correlations split into these simple second-order correlations (Wick theorem \cite{wick})
\begin{align}
&\langle \phi(x_1)\cdots\phi(x_n)\rangle=\nonumber\\
&2^{-n/2}\sum_{\sigma}\prod_{j=1}^{n/2}\langle\phi(x_{\sigma(j)})\phi(x_{\sigma(n/2+j)})\rangle
\end{align}
for even $n$ while $0$ for odd $n$, summing over all permutations. For the fermionic field, one has to include also
the permutation sign $\mathrm{sgn}\, \sigma$. In the case of linear perturbation, we often use  the identity
$
\int da e^{-2\lambda a^2+ba}=(\pi/2\lambda)^{1/2}e^{b^2/8\lambda}
$.
Applying functional derivative
\begin{align}
&\delta(x-y)=\frac{\delta\phi(y)}{\delta\phi(x)}=\left\langle\frac{\delta\phi(y)}{\delta\phi(x)}\right\rangle\nonumber\\
&=
Z\int \mathcal D\phi \frac{\delta\phi(y)}{\delta\phi(x)}\exp\int i\mathcal L(z)dz
\end{align}
 and integrating by parts
\begin{align}
&\delta(x-y)Z^{-1}=
\int \mathcal D\phi \phi(y)\int\frac{\delta\mathcal L(w)}{i\delta\phi(x)}dw\exp\int i\mathcal L(z)dz\nonumber\\
&
=\int \mathcal D\phi i\phi(y)(\partial\cdot\partial\phi(x)+m^2\phi(x))\exp\int i\mathcal L(z)dz
\end{align}
we get
\begin{equation}
(g^{\mu\nu}\partial_\mu \partial_\nu+m^2)\langle \phi(x)\phi(y)\rangle=\delta(x-y)/i
\end{equation}
with the derivative over $x$.
The equation can be solved by Fourier transform in $\boldsymbol x$ and $\boldsymbol y$, i.e.
\begin{equation}
\phi(x)=\int\frac{d\boldsymbol k}{(2\pi)^{D/2}}e^{i\boldsymbol k\cdot \boldsymbol x}\phi(x^0,\boldsymbol k)
\end{equation}
with the standard scalar product $\boldsymbol k\cdot\boldsymbol x=\sum_{j=1}^Dk^jx^j$.
We obtain
\begin{align}
&(\partial_0\partial_0+\boldsymbol k\cdot\boldsymbol k+m^2)\langle \phi(x^0,\boldsymbol k)\phi(y^0,\boldsymbol q)\rangle\nonumber\\
&=\delta(x^0-y^0)\delta(\boldsymbol k-\boldsymbol q)/i
\end{align}
with the standard $\delta(\boldsymbol k-\boldsymbol q)=\delta(k^1-q^1)\cdots\delta(k^D-q^D)$.
This is a simple linear equation with the solution
\begin{equation}
\langle \phi(x^0,\boldsymbol k)\phi(y^0,\boldsymbol q)\rangle=\sum_\pm \pm\delta(\boldsymbol k-\boldsymbol q)
\frac{e^{\mp i|x^0-y^0|E_k}}{1-e^{\mp \beta E_k}}
\end{equation}
with $E_k=\sqrt{\boldsymbol k\cdot\boldsymbol k+m^2}$ and
\begin{equation}
|x^0-y^0|=\left\{\begin{array}{ll}
x^0-y^0&\mbox{ if }s_x>s_y\\
y^0-x^0&\mbox{ otherwise }\end{array}\right.
\end{equation}
In the zero-temperature limit $\beta\to 0$ we get
\begin{equation}
\langle \phi(x^0,\boldsymbol k)\phi(y^0,\boldsymbol q)\rangle=\delta(\boldsymbol k-\boldsymbol q)
e^{-i|x^0-y^0|E_k}.
\end{equation}
Equivalently
\begin{equation}
\langle \phi(x)\phi(y)\rangle=\sum_\pm \int \frac{\pm d\boldsymbol k}{(2\pi)^D 2E_k}e^{i\boldsymbol k\cdot(\boldsymbol x-\boldsymbol y)}
\frac{e^{\mp i|x^0-y^0|E_k}}{1-e^{\mp\beta E_k}}
\end{equation}
or in the zero-temperature limit
\begin{equation}
\langle \phi(x)\phi(y)\rangle=\int \frac{d\boldsymbol k}{2E_k(2\pi)^D}e^{i\boldsymbol k\cdot(\boldsymbol x-\boldsymbol y)}e^{-i|x^0-y^0|E_k}.
\end{equation}
The special correlations read
\begin{align}
&S(x)=\sum_\pm \int \frac{\pm d\boldsymbol k}{(2\pi)^D 2E_k}e^{i\boldsymbol k\cdot\boldsymbol x}
\frac{e^{\mp i|x^0|E_k}}{1-e^{\mp\beta E_k}},\nonumber\\
&
B(x)=\sum_\pm \int \frac{\pm d\boldsymbol k}{(2\pi)^D 2E_k}e^{i\boldsymbol k\cdot\boldsymbol x}
\frac{e^{\pm ix^0E_k}}{1-e^{\mp\beta E_k}},\nonumber\\
&
C(x)=\int \frac{d\boldsymbol k}{(2\pi)^D 2E_k}e^{i\boldsymbol k\cdot\boldsymbol x}\frac{\cos(x^0E_k)}{\tanh(\beta E_k/2)},
\end{align}
with $|x^0|$ reduced to usual absolute value
and zero-temperature limits
\begin{align}
&S(x)=\int \frac{d\boldsymbol k}{2E_k(2\pi)^D}e^{i\boldsymbol k\cdot\boldsymbol x}e^{-i|x^0|E_k}\nonumber\\
&
=\int\frac{idk}{(2\pi)^{D+1}}\frac{e^{ik\cdot (x-y)}}{k\cdot k-m^2+i\epsilon},
\nonumber\\
&B(x)=\int \frac{d\boldsymbol k}{2E_k(2\pi)^D}e^{i\boldsymbol k\cdot\boldsymbol x}e^{ix^0E_k}\nonumber\\
&
= \int \frac{ dk}{(2\pi)^{D}}\delta(k\cdot k-m^2)e^{ik\cdot x}\theta(k^0),\nonumber\\
&C(x)=\int \frac{d\boldsymbol k}{(2\pi)^D 2E_k}e^{i\boldsymbol k\cdot\boldsymbol x}\cos(x^0E_k)\nonumber\\
&
=\int \frac{ dk}{2(2\pi)^{D}}\delta(k\cdot k-m^2)e^{ik\cdot x},\nonumber\\
&G(x)=\int \frac{d\boldsymbol k}{iE_k(2\pi)^D}e^{i\boldsymbol k\cdot\boldsymbol x}\sin(x^0E_k)\nonumber\\
&=\int\frac{idk}{(2\pi)^{D+1}}\frac{e^{ik\cdot x}}{k_+\cdot k_+-m^2}.\label{zzz}
\end{align}
The causal Green function is independent of temperature and defined only for $x^0\geq 0$
while $G=0$ for $x\cdot x<0$, where $k^0_+=k^0-i\epsilon$ ($\epsilon\to 0_+$ as previously is
necessary to make the integrals well defined).

 Convenient substitution $\boldsymbol k= k\boldsymbol n$ where $\boldsymbol n$ is a unit vector
and $k=m\sinh\eta>0$. Then $E_k=m\cosh\eta$ and $d\boldsymbol k/E_k=d\boldsymbol n (m\sinh \eta)^{D-1}d\eta$ (in $D=1$ we have $\sum_{\boldsymbol n=\pm 1}$ instead of $\int d\boldsymbol n$).
Then
\begin{align}
&\langle \phi(x)\phi(0)\rangle=\int \frac{(m\sinh \eta)^{D-1}d \eta d\boldsymbol u}{2(2\pi)^D}\times\nonumber\\
&e^{i m\sinh\eta\boldsymbol u\cdot\boldsymbol x}e^{-i|x^0|m\cosh\eta}
\end{align}
For $x^{1\dots D}=0$ we have
\begin{equation}
\langle \phi(x)\phi(0)\rangle=\int \frac{(m\sinh \eta)^{D-1}d \eta d\boldsymbol u}{2(2\pi)^D}e^{-i|x^0|m\cosh\eta}.
\end{equation}
The integral $\int d\boldsymbol u=S_D$ is the surface of the $D-1$ unit sphere (embedded in $D$ dimensions). 
In particular $S_1=2$, $S_2=2\pi$, $S_3=4\pi$, or in general $S_D=2\pi^{D/2}/\Gamma(D/2)$ ($\Gamma$ - Euler Gamma function, here $\Gamma(1/2)=\pi^{1/2}$,
$\Gamma(1)=1$ and $\Gamma(z+1)=z\Gamma(z)$). Substituting $w=\cosh\eta$ we get
\begin{align}
&\int \frac{m^{D-1}(w^2-1)^{D/2-1}dw d\boldsymbol u}{2^{D}\pi^{D/2}\Gamma(D/2)}e^{-i|x^0|mw}=\nonumber\\
&\frac{m^{(D-1)/2}K_{(D-1)/2}(i|x^0|m)}{(2\pi)^{(D+1)/2}(i|x^0|)^{(D-1)/2}}.
\end{align}

By Lorentz invariance and analyticity we can write in general
\begin{align}
&\langle \phi(x)\phi(0)\rangle=\nonumber\\
&\frac{m^{(D-1)/2}K_{(D-1)/2}(m\sqrt{-x\cdot x})}{(2\pi)^{(D+1)/2}(-x\cdot x)^{(D-1)/4}}
\end{align}
with the complex square root defined so that the real part is positive. The divergence at $x=0$ is removable because we can make an
 infinitesimal shift in the imaginary direction.

We shall list the special cases in the zero-temperature limit \cite{prop}.
%based on \url{https://arxiv.org/pdf/0811.1261.pdf} and \url{http://scipp.ucsc.edu/~haber/webpage/DeltaF.pdf}.
Case $D=1$:
\begin{align}
&C(x)=\left\{\begin{array}{ll}
K_0(m\sqrt{-x\cdot x})/2\pi&\mbox{ for }x\cdot x<0,\\
-Y_0(m\sqrt{x\cdot x})/4&\mbox{ for }x\cdot x>0,\end{array}\right.\nonumber\\
&
G(x,y)=-iJ_0(m\sqrt{x\cdot x})/2\mbox{ for }x\cdot x>0\mbox{ and }x^0>0.
\end{align}
Case $D=2$. 
Taking into account that $K_{1/2}(z)=e^{-z}\sqrt{\pi/2z}$ we can write
\begin{equation}
\langle \phi(x)\phi(0)\rangle=\frac{\exp(-\sqrt{-x\cdot x}m)}{4\pi\sqrt{-x\cdot x}}
\end{equation}
or, in particular,
\begin{align}
&C(x)=\frac{C'(x)}{4\pi\sqrt{|x\cdot x|}},\nonumber\\
&C'(x)=\left\{\begin{array}{ll}
\exp(-m\sqrt{-x\cdot x})&\mbox{ for }x\cdot x<0\\
-\sin(m\sqrt{x\cdot x})&\mbox{ for }x\cdot x>0\end{array}\right.,\nonumber\\
&G(x)=\frac{\cos(\sqrt{x\cdot x}m)}{2\pi i\sqrt{x\cdot x}},
\end{align}
for $x^0>0$ and $x\cdot x>0$.
Case $D=3$:
\begin{align}
&\langle \phi(x)\phi(0)\rangle=\frac{m}{4\pi^2\sqrt{-x\cdot x}}K_1(m\sqrt{-x\cdot x}),\nonumber\\
&C(x)=\frac{mC'(x)}{8\pi^2\sqrt{|x\cdot x|}},\nonumber\\
&C'(x)=\left\{\begin{array}{ll}
2K_1(m\sqrt{-x\cdot x})&\mbox{ for }x\cdot x<0,\\
\pi Y_1(m\sqrt{x\cdot x})&\mbox{ for }x\cdot x>0,\end{array}\right.\nonumber\\
&G(x)=\frac{imJ_1(\sqrt{x\cdot x}m)}{4\pi\sqrt{x\cdot x}}+\frac{\delta( x\cdot x)}{2\pi i},
\end{align}
with only the last term in the $m\to 0$ case.

The $m\to 0$ case. By expansion of Bessel functions, we have for $D=1$
\begin{equation}
-2\pi\langle \phi(x)\phi(0)\rangle\to \ln(-x\cdot x)/2+\ln(m/2)+\gamma
\end{equation}
with the Euler-Mascheroni constant $\gamma$ and $G(x)\to -i/2$ (for $x^0>0$ and $x\cdot x>0$). 
One encounters infrared divergence of correlation at small $\boldsymbol k$ and $m\to 0$.
For $D>1$ we have

\begin{equation}
\langle \phi(x)\phi(0)\rangle\to \frac{\Gamma((D-1)/2)}{4\pi^{(D+1)/2}(-x\cdot x)^{(D-1)/2}}.
\end{equation}
For $D=2$,
\begin{align}
&G(x)\to 1/2\pi i\sqrt{x\cdot x},\nonumber\\
&
C(x)\to \theta(-x\cdot x)/4\pi\sqrt{-x\cdot x}.
\end{align}
For $D=3$,
\begin{align}
&G(x)\to \delta(x\cdot x)/2\pi i,\nonumber\\
&C(x)\to -1/4\pi^2 x\cdot x.
\end{align}
Since $C$ diverges at $x\cdot x\to 0$ (also for $m>0$)  we have to calculate it as Cauchy principal value.

\section{VACUUM FLUCTUATIONS}
\label{appc}

\begin{figure}
\includegraphics[scale=.5]{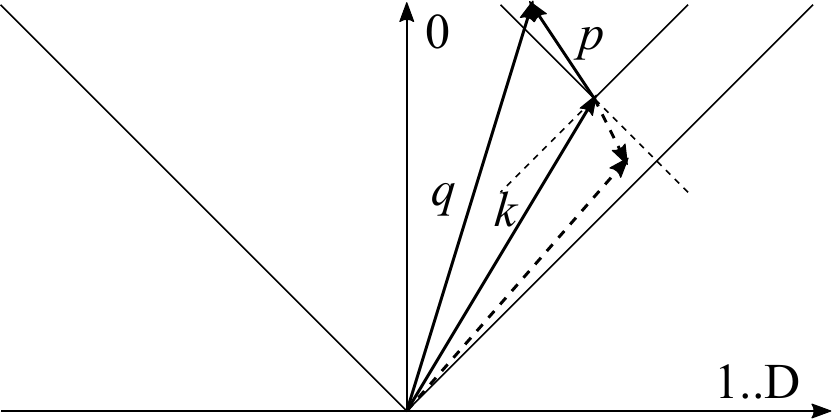}
\caption{Relation between $k$, $p$ and $q$ showing that timelike $p=q-k$ is not necessarily forward, even if $k$ and $q$ are.}
\label{kpq}
\end{figure}

Here we present the details of the 
calculation of (\ref{aa0}). We begin with
\begin{align}
&\langle a^2\rangle_\bullet=
\int dkdp\tilde f(k+p)\tilde f(-k-p)\theta(k^0)\theta(p^0)\nonumber\\
&\times 8\pi^2\delta(k\cdot k-m^2)\delta(p\cdot p-m^2).
\end{align}
Note that $k$ and $p$ are forward, i.e. $k\cdot k,p\cdot p>0$ and $k^0,p^0>0$ so $q=k+p$ is also forward. 
On the other hand, forward $q$ and timelike $k,p$ are not sufficient to keep  both $k$ and $p$ forward, see Fig. \ref{kpq}.

Replacing $k+p=q$ and shifting $p$, we get
\begin{align}
&\int dqdp\tilde f(q)\tilde f(-q)\theta(q^0/2+p^0)\theta(q^0/2-p^0)8\pi^2\times\nonumber\\
&\delta((q/2+p)\cdot(q/2+p)-m^2)\times\nonumber\\
&\delta((q/2-p)\cdot(q/2-p)-m^2).
\end{align}
Note that
\begin{align}
&\delta((q/2+p)\cdot(q/2+p)-m^2)\times\nonumber\\
&\delta((q/2+p)\cdot(q/2+p)-m^2)\nonumber\\
&=\delta(2q\cdot p)\delta(q\cdot q/4+p\cdot p-m^2),
\end{align}
which gives fixings $q\cdot p=0$ and $q\cdot q/4+p\cdot p=m^2$.

We need to calculate $W(q)$ equal to
\begin{equation}
\int dp\theta(|q^0|/2-|p^0|)\delta(2q\cdot p)\delta(q\cdot q/4+p\cdot p-m^2)/2
\end{equation}
but due to Lorentz invariance, we need to do it only for $q^{1\dots D}=0$ and $q^0>0$, which is the easiest. Then $p^0=0$ and by substitution 
$|\boldsymbol p|=\tilde p$ it reduces to
\begin{equation}
\int \tilde{p}^{D-1}S_Dd\tilde{p}\delta((q^0)^2/4-\tilde{p}^2-m^2)/4q^0.
\end{equation}
Finally putting $p=\sqrt{(q^0)^2/4-m^2}$ and replacing $q^0\to \sqrt{q\cdot q}$, we get (\ref{www}).

\section{PLAVE WAVE GENERATIONS}
\label{appd}

We shall derive the causal Green function in the limit of oscillating perturbation that generates a plane wave.
We define
\begin{align}
&G_g(x)=\int dy G(x,y)g(y)=nonumber\\
&=\int_{y^0<x^0} dy \frac{d\boldsymbol k}{iE_k(2\pi)^D}e^{i\boldsymbol k\cdot(\boldsymbol x-\boldsymbol y)}\sin((x^0-y^0)E_k)g(y).
\end{align}
We first integrate over $y^{2,3,...}$ while substituting $y^0=x^0-t$, and $y^1=y-L$ getting
\begin{align}
&\int_{t>0} dtdy \frac{dk}{iE_k(2\pi)}e^{ik(x^1-y)}e^{ikL}\nonumber\\
&\times\sin(tE_k)
(e^{iE_p(x^0-t)}h(y)+e^{-iE_p(x^0-t)}h^\ast(y))=\nonumber\\
&\times\int_{t>0} dtdy \frac{dk}{E_k(4\pi)}e^{ik(x^1-y)}e^{ikL}\nonumber\\
&[(e^{itE_k+iE_p(x^0-t)}-e^{-itE_k+iE_p(x^0-t)})h(y)\nonumber\\
&
+(e^{itE_k-iE_p(x^0-t)}-e^{-itE_k-iE_q(x^0-t)})h^\ast(y)]
\end{align}
with $k$ being a one-dimensional  variable and $E_k=\sqrt{k^2+m^2}$. We integrate over $t$ with damping factor $e^{-0_+ t}$.
Then
\begin{align}
&G_g(x)=\int\frac{dk}{E_k(4\pi)}e^{ik(L+x^1)}\nonumber\\
&\times\left[\left(\frac{e^{iE_px^0}}{0_+-iE_k+iE_p}-\frac{e^{iE_px^0}}{0_++iE_k+iE_p}\right)\tilde{h}(k)\right.\nonumber\\
&\left.
+\left(\frac{e^{-iE_px^0}}{0_+-iE_k-iE_p}-\frac{e^{-iE_px^0}}{0_+ +iE_k-iE_p}\right)\tilde{h}^\ast(-k)\right]
\end{align}
with $\tilde{h}(k)=\int dy h(y)e^{-iky}$. For large $L$ the factor $e^{ikL}$ is quickly oscillating while the other functions vary slowly.
Exceptions are the first and last denominators at $E_k\sim E_p$ as they diverge. Only in these cases, we make approximations
$
E_k-E_p\simeq (|k|-|p|)|p|/E_p
$
for $|p|=\sqrt{E_p^2-m^2}$. The integral concentrates only near two peaks at $k=\pm |q|$ so we can calculate
\begin{align}
&G_g(x)\to\sum_{\pm}\int \frac{ dk}{4\pi E_p}e^{ik(L+x^1)}\times\nonumber\\
&\left[\frac{e^{iE_px^0}\tilde{h}(\pm |p|)}{0_+ +i|p|(|p|\mp k)/E_p}-
\frac{e^{-iE_px^0}\tilde{h}^\ast(\mp |p|)}{0_+ -i|p|(|p|\mp k)/E_p}\right]\nonumber\\
&=\sum_{\pm}\int \frac{ dk}{4\pi|p|}e^{ik(L+x^1)}\times\nonumber\\
&\left[\frac{e^{iE_px^0}\tilde{h}(\pm |p|)}{0_+ +i(|p|\mp k)}-
\frac{e^{-iE_px^0}\tilde{h}^\ast(\mp |p|)}{0_+ -i(|p|\mp k)}\right].
\end{align}
The integral over $k$ can now be calculated using residues, giving
\begin{align}
&2|p|G_g(x)\to\nonumber\\
&e^{iE_px^0-i|p|x^1}\tilde{h}(- |p|)-e^{i|p|x^1-iE_px^0}\tilde{h}^\ast(+|p|)
\end{align}
equivalent to (\ref{gga}).

\section{CONTACT TERM PROBLEM IN ENERGY CORRELATIONS}
\label{appe}

 For the normalization $\langle 1\rangle_0=1$ (unitarity) to hold, we expect 
$
\langle T^{\mu\nu}_q(x)T^{\xi\eta}_q(y)\rangle
$
to vanish
because any quantity $A_q(x)=A_+(x)-A_-(x)$ should cancel the correlation if $x^0$ is the \emph{latest} time.
For $\mu=\nu=\xi=\eta=0$ the above expression will contain the term
\begin{equation}
 (\partial^0_x\partial^0_yG(x,y))(\partial^0_x\partial^0_yG(y,x)),
\end{equation}
which indeed is $0$ for both $x^0>y^0$ and $y^0-x^0$. However, something strange happens at $x^0=y^0$. Then taking the definitions of $G$
we get
\begin{equation}
\partial^0_xG(x,y)=\int \frac{d\boldsymbol k}{i(2\pi)^D}e^{i\boldsymbol k\cdot(\boldsymbol x-\boldsymbol y)}\cos((x^0-y^0)E_k)
\end{equation}
for $x^0>y^0$. and $0$ for $x^0<y^0$. Unfortunately, there is a discontinuity at $x^0=y^0$ which gives
$
 \partial^0_x\partial^0_yG(x,y)=i\delta(x-y)
$
As a result, we get $\delta^2(x-y)=\delta(0)\delta(x-y)$, with $\delta(0)$ undefined, formally divergent to $+\infty$.
The problem persists at higher (arbitrary) order correlations since we get e.g.
\begin{align}
&\delta(x-y)\delta(y-z)\delta(z-w)\delta(w-x)=\nonumber\\
&\delta(x-y)\delta(x-y)\delta(y-z)\delta(z-w)\delta(0)
\end{align}
again diverging.
To cure the problem, we need an auxiliary independent renormalizing field $\phi_M$,  
but with a large mass $M\to\infty$, 
and construct $T^{\mu\nu}_M(x)$ by replacing $m\to M$ and $\phi\to\phi_M$. Unfortunately, $\phi_M$ must be anticommuting (Grassmann, fermionic).
More precisely, there are two fields $\phi_M(x)$ and $\phi^\ast_M(y)$ but they anticommute, i.e., $AB=-BA$ for $A,B=\phi_M(x,y),\phi_M^\ast(x,y)$.  
The field is called a ghost because it is not physical \cite{fadeev}.
The renormalizing Lagrangian density reads
\begin{equation}
\mathcal L_M(x)=\partial\phi^\ast_M(x)\cdot\partial\phi_M(x)-m^2\phi^\ast(x)\phi(x)
\end{equation}
with the energy-momentum tensor
\begin{align}
&T^{\mu\nu}_M(x)=\partial^\mu\phi^\ast_M\partial^\nu\phi_M+\partial^\nu\phi^\ast_M\partial^\mu\phi_M\nonumber\\
&-g^{\mu\nu}(g^{\sigma\tau}\partial_\sigma\phi^\ast_M\partial_\tau\phi_M-M^2\phi^\ast_M\phi_M).
\end{align}
The basic correlations read $\langle\phi_M(x)\phi_M(y)\rangle=0$ and
\begin{align}
&\langle \phi_M(x)\phi^\ast_M(y)\rangle=\nonumber\\
&\int \frac{d\boldsymbol k}{(2\pi)^D 2E_k}e^{i\boldsymbol k\cdot(\boldsymbol x-\boldsymbol y)}
\left(\frac{e^{-i|x^0-y^0|E_k}}{1+e^{-\beta E_k}}-\frac{e^{i|x^0-y^0|E_k}}{1+e^{\beta E_k}}\right)
\end{align}
with the zero-temperature limit
\begin{equation}
\langle \phi_M(x)\phi^\ast_M(y)\rangle=\int \frac{d\boldsymbol k}{(2\pi)^D 2E_k}e^{i\boldsymbol k\cdot(\boldsymbol x-\boldsymbol y)}e^{-i|x^0-y^0|E_k}.
\end{equation} 
The Wick decomposition includes now the sign of the permutation
\begin{align}
&\langle \phi^\ast_M(x_1)\phi_M(y_1)\cdots\phi^\ast_M(x_n)\phi_M(y_n)\rangle=\nonumber\\
&\sum_\sigma \mathrm{sgn}\,\sigma\langle\phi^\ast_M(x_{\sigma(1)})\phi_M(y_1)\rangle\cdots\langle\phi^\ast_M(x_{\sigma(n)})\phi_M(y_n)\rangle.
\end{align}
Now, because of the opposite sign of the permutation we get
\begin{equation}
\langle T^{\mu\nu}_{Mq}(x)T^{\xi\eta}_{Mq}(y)\rangle=-\langle T^{\mu\nu}_{q}(x)T^{\xi\eta}_{q}(y)\rangle
\end{equation}
so let us modify $T^{\mu\nu}$ by
\begin{equation}
T^{\mu\nu}(x)\to T^{\mu\nu}(x)+T^{\mu\nu}_{M}(x)
\end{equation}
in our Kraus operator. Then what we get is
\begin{equation}
\langle T^{\mu\nu}_q(x)T^{\xi\eta}_q(y)\rangle=0.
\end{equation}
Note that finite $M$ would also add some correction to terms containing $T^{\mu\nu}_c$ (which do not spoil unitarity, though) so we keep the limit $M\to\infty$.

\section{ENEGRY MOMENTUM FLUCTUATIONS}
\label{appf}

We define
\begin{equation}
\tilde f_{\mu\nu}(k)=\int dx e^{ik\cdot x}f_{\mu\nu}(x)/(2\pi)^{D+1}
\end{equation}
and note that
\begin{align}
&\langle T^{\mu\nu}_c(x)T^{\xi\eta}_c(y)\rangle=\nonumber\\
&\langle T^{\mu\nu}_c(x)T^{\xi\eta}_c(y)\rangle-\langle T^{\mu\nu}_q(x)T^{\xi\eta}_q(y)\rangle/4
\nonumber\\
&=\langle(T^{\mu\nu}(x_+)T^{\xi\eta}_c(y_-)+T^{\mu\nu}(x_-)T^{\xi\eta}_c(y_+))\rangle/2.
\end{align}
Using $B(x,y)$ and $B(y,x)$ we get
\begin{align}
&\langle a^2\rangle_\bullet=
\int dkdp\tilde f_{\mu\nu}(k+p)\tilde f_{\xi\eta}(-k-p)\nonumber\\
&\times(k^\mu p^\nu-g^{\mu\nu}(k\cdot p+m^2)/2)\nonumber\\
&\times(k^\xi p^\eta-g^{\xi\eta}(k\cdot p+m^2)/2)\nonumber\\
&\times 8\pi^2\theta(k^0)\theta(p^0)\delta(k\cdot k-m^2)\delta(p\cdot p-m^2).
\end{align}
Replacing $k+p=q$ and shifting $p$, we get
\begin{equation}
\langle a^2\rangle_\bullet=\int dq\tilde f_{\mu\nu}(q)\tilde f_{\xi\eta}(-q)X^{\mu\nu\xi\eta}(q)
\end{equation}
with
\begin{align}
&X^{\mu\nu\xi\eta}(q)=\int dp\theta(|q^0|/2-|p^0|)\delta(2q\cdot p)4\pi^2\times\nonumber\\
&\delta(q\cdot q/4+p\cdot p-m^2)\times
\nonumber\\
&(q^\mu q^\nu/4 -p^\mu p^\nu-g^{\mu\nu}q\cdot q/4)\times\nonumber\\
&(q^\xi q^\eta/4 -p^\xi p^\eta-g^{\xi\eta}q\cdot q/4).
\end{align}
We can observe that (a) $X$ is $0$ at $q\cdot q<0$, (b) $X$ is symmetric under interchanging
$\mu\leftrightarrow \nu$ or $\xi\leftrightarrow \eta$ or $\mu\nu\leftrightarrow \xi\eta$, (c) $X$ satisfies $q_\mu X^{\mu\nu\xi\eta}(q)=0$ (Ward identity or energy conservation) (d) is Lorentz covariant (there is no preferred frame).
Even more, if we take $q^{1\cdots D}=0$, then the constraints lead to $p^0=0$ and $|\boldsymbol p|^2=(q^0)^2/4=m^2$ so we have bound $q\cdot q> 4m^2$.
Therefore the expected form of $X$ is
\begin{align}
&X^{\mu\nu\xi\eta}=\theta(q\cdot q-4m^2)4\pi^2\nonumber\\
&\times[
P'(q^\mu q^\nu -(q\cdot q) g^{\mu\nu})(q^\xi q^\eta-(q\cdot q)g^{\xi\eta})
\nonumber\\
&+R'[(q^\mu q^\eta -(q\cdot q) g^{\mu\eta})(q^\xi q^\nu-(q\cdot q)g^{\xi\nu})\nonumber\\
&+
(q^\mu q^\xi -(q\cdot q) g^{\mu\xi})(q^\nu q^\eta-(q\cdot q)g^{\nu\eta})]].
\end{align}
We only need to find two functions $P'(q\cdot q)$ and $R'(q\cdot q)$, which is the simplest by contraction
\begin{equation}
g_{\mu\nu}g_{\xi\eta}X^{\mu\nu\xi\eta}=(q\cdot q)^2D(DP'+2R')
\end{equation}
and
\begin{equation}
g_{\mu\xi}g_{\nu\eta}X^{\mu\nu\xi\eta}=(q\cdot q)^2D[P'+R'(D+1)]
\end{equation}
giving
\begin{align}
&(D-1)(D+2)D(q\cdot q)^2P'=\nonumber\\
&(D+1)g_{\mu\nu}g_{\xi\eta}X^{\mu\nu\xi\eta}-2g_{\mu\xi}g_{\nu\eta}X^{\mu\nu\xi\eta},
\nonumber\\
&(D-1)(D+2)D(q\cdot q)^2R'=\nonumber\\
&Dg_{\mu\xi}g_{\nu\eta}X^{\mu\nu\xi\eta}-g_{\mu\nu}g_{\xi\eta}X^{\mu\nu\xi\eta}.
\end{align}
At $D=1$ both equations give $P'+2R'$ but it is not a problem, as the $P'$ and $R'$ terms are actually the same.
It is clear from the  general rule $X^{0\nu\xi\eta}=X^{1\nu\xi\eta}q^1/q^0$ and similarly for all indices
so  $X$ is fixed by just one term $X^{1111}$ (not true at $D>1$).

We can express the contraction by previous quantities
\begin{align}
&g_{\mu\nu}g_{\xi\eta}X^{\mu\nu\xi\eta}=\nonumber\\
&\int dp\theta(|q^0|/2-|p^0|)\delta(2q\cdot p)\delta(q\cdot q/4+p\cdot p-m^2)
\nonumber\\
&\times
(q\cdot q/4 -p\cdot p-(D+1)q\cdot q/4)^2
\nonumber\\
&
=((D-1)q\cdot q/4 +m^2)^2W(q)
\end{align}
and
\begin{align}
&g_{\mu\xi}g_{\nu\eta}X^{\mu\nu\xi\eta}=\nonumber\\
&\int dp\theta(|q^0|/2-|p^0|)\delta(2q\cdot p)\delta(q\cdot q/4+p\cdot p-m^2)
\nonumber\\
&\times
((q\cdot q)^2/16+(p\cdot p)^2+(D+1)(q\cdot q)^2/16\nonumber\\
&-(q\cdot p)^2/2-(q\cdot q)^2/8+(p\cdot p)(q\cdot q)/2)
\nonumber\\
&
=((q\cdot q)^2(D-1)/16+m^4)W(q)
\end{align}
for $W$ given by (\ref{www})
so that
\begin{align}
&(q\cdot q)^2(D+2)DP'=\nonumber\\
&(m^4+(D+1)m^2q\cdot q/2+(q\cdot q)^2(D^2-3)/16)W(q),
\nonumber\\
&(q\cdot q)^2(D+2)DR'=\nonumber\\
&(m^4-m^2q\cdot q/2+(q\cdot q)^2/16)W(q)
\end{align}
with $P'=PW$, $R'=RW$ in (\ref{ttt}).

In the case $D=1$ we can set $R'=0$ and
\begin{equation}
P'=\pi^2 m^4(q\cdot q)^{-5/2}(q\cdot q/4-m^2)^{-1/2}
\end{equation}
In the limit $m\to 0$ also $P'\to 0$ but only at $q\cdot q> 0$. The limit $q\cdot q\to 0$ has to be considered separately
Note that the final integral 
\begin{align}
&\int dq \theta(q\cdot q/4-m^2)m^4(q\cdot q)^{-5/2}\times\nonumber\\
&(q\cdot q/4-m^2)^{-1/2}|(q^\mu q^\nu -g^{\mu\nu}(q\cdot q))\tilde{f}_{\mu\nu}(q)|^2\pi^2
\end{align}
can be transformed using the change of variables $q^0=m\sqrt{w}\cosh u$, $q^1=m\sqrt{w}\sinh u$ into
\begin{align}
&\int_{4}^\infty dw\int du w^{-5/2}(w/4-1)^{-1/2}m^4w^2\times\nonumber\\
&|(U^\mu U^\nu -g^{\mu\nu})\tilde{f}_{\mu\nu}(m\sqrt{w}U)|^2\pi^2
\end{align}
with $U^0=\cosh u$, $U^1=\sinh u$. Suppose $\tilde{f}(q)$ is a regular function that decays to $0$ if $q\to \infty$.
Then the limit $m\to 0$ reduces the integral essentially to the lines of light, i.e. $q^0=\pm q^1$, for $|u|\gg 1$.
We make then the approximation $U=e^u U_\pm/2$ with $\lambda=m\sqrt{w}e^u$ and $U_\pm=(1,\pm 1)$. Then we get
\begin{equation}
\langle a^2\rangle_0=\sum_\pm\int_0^\infty d\lambda \lambda^3|U_\pm^\mu U_\pm^\nu \tilde{f}_{\mu\nu}(\lambda U_\pm)|^2\pi^2/3\cdot 2^5
\end{equation}
since
\begin{equation}
\int_{4}^\infty dw w^{-5/2}(w/4-1)^{-1/2}=1/6.
\end{equation}

\end{document}